\documentclass[reprint,amsmath,amssymb,aps,floatfix,pre,superscriptaddress]{revtex4-1}

\usepackage{graphicx}
\usepackage{dcolumn}
\usepackage{bm}

\begin{document}

\preprint{}

\title{Static and dynamic shear viscosity of a single layer complex plasma}

\author{Peter Hartmann}
\affiliation{Research Institute for Solid State Physics and Optics of the Hungarian Academy of Sciences, P.O.B. 49, H-1525 Budapest, Hungary}
\affiliation{Center for Astrophysics, Space Physics and Engineering Research (CASPER), One Bear Place 97310, Baylor University, Waco, TX 76798, USA}
\author{M\'at\'e Csaba S\'andor}
\author{Anik\'o Kov\'acs}
\author{Zolt\'an Donk\'o}
\affiliation{Research Institute for Solid State Physics and Optics of the Hungarian Academy of Sciences, P.O.B. 49, H-1525 Budapest, Hungary}

\date{\today}

\begin{abstract}
The static and dynamic (complex) shear viscosity of a single layer dusty plasma is measured by applying, respectively, a stationary and a periodically modulated shear stress, induced by the light pressure of manipulating laser beams. Under static conditions we observe a decrease of the viscosity with increasing shear rate, the so-called shear-thinning behavior. Under oscillating shear both the magnitude and the ratio of the dissipative and elastic contributions to the complex viscosity show strong frequency dependence, as the system changes from viscous to elastic in nature with increasing excitation frequency. Accompanying molecular dynamics simulations explain and support the experimental observations.
\end{abstract}


\maketitle

\section{Introduction}

Transport properties of strongly coupled complex (or dusty) plasmas,  have attracted considerable attention during the past decade (see e.g. \cite{Morfill2009,*Bonitz2010}). The interaction of the dust particles in these systems may be well approximated by the Debye-H\"uckel (or Yukawa) model potential $\Phi(r) = Q \exp[-r/\lambda_D]~/~4 \pi\varepsilon_0 r$, where $Q$ is the dust particle charge, $\lambda_D$ is the Debye screening length. Equilibrium Yukawa systems can be fully characterized by two dimensionless quantities: the Coulomb coupling parameter $\Gamma = Q^2 / (4 \pi\varepsilon_0 a k_\text{B} T)$, where $T$ is the temperature, and the screening parameter $\kappa=a/\lambda_D$.  

Viscosity, the measure of the plastic response of matter (primarily liquids and soft matter) to applied forces, is a central quantity in rheology. For a Newtonian fluid a constant shear viscosity $\eta$ relates the shear stress $\sigma$ to the shear rate $\dot{\gamma}$ = $\partial v_x/\partial y$ (velocity gradient) as $\sigma = -\eta \dot{\gamma} = -\eta (\partial v_x/\partial y)$. 

Continuum hydrodynamics successfully uses the concept of viscosity, usually as input parameter, as an intrinsic property of the material under investigation, in the Navier-Stokes equation. One has to note, however, that the Newtonian concept of viscosity is applicable only (i) at small shear rates, (ii) long length scales, and (iii) at low frequencies. In many physical systems, these conditions are clearly violated \cite{Schowalter}. 

Regarding complex plasmas, the first experiment on a single layer (2D) dust cloud with control over the applied shear, via adjusting the power of a manipulating laser beam, was carried out by the group of Lin I in 2001 \cite{LinI2001}. Similarly, an experiment making use of a single shearing laser beam was reported in \cite{Gavrikov05}. In these experiments a sheared velocity profile was created around the beam. In the experiment reported in \cite{NosenkoPRL04} two displaced, parallel counter-propagating laser beams were used to realize a planar Couette configuration in a 2D dusty plasma layer. In another experiment the non-Newtonian behavior of a 3D complex plasma in the liquid state was identified by Ivlev {\it et al.} \cite{IvlevPRL07}. Motivated by these pioneering studies, more recently, detailed dusty plasma experiments demonstrated the presence of viscoelastic response \cite{LinI2007} and  revealed the wave-number dependence of the viscosity in 2D \cite{GoreePRL10}. Experiments in the crystalline phase have identified the slipping of individual crystal lines to be the primary mechanism for relaxing an applied shear stress \cite{Samsonov2011}.

Complementing the experimental efforts, the self-diffusion \cite{Ohta,Ramazanov06,Torben08,*Torben09,Feng10}, shear viscosity \cite{Saigo,Salin1,*Salin2,Murillo,DonkoV,Ramazanov} and thermal conductivity coefficients \cite{Salin1,*Salin2,DonkoT} have been derived in a number of simulation studies, both for three-dimensional (3D) and two-dimensional (2D) settings. In \cite{DonkoPRL06,*DonkoMPL07}, besides calculations of the ``equilibrium'' ($\dot{\gamma} \rightarrow 0$) static viscosity, predictions for the shear-thinning effect (typical for complex molecular liquids) were given at high shear rates. The frequency dependence of the complex shear viscosity, which combines the dissipative and the elastic components of the complex response of matter to oscillating shear stress as $\eta(\omega) = \eta^\prime(\omega) - i\eta^{\prime\prime}(\omega)$, was computed for 3D Yukawa liquids in \cite{DonkoPRE10}. Fundamental questions about the existence or nonexistence of well defined transport coefficients in 2D were addressed in \cite{DonkoPRE09}.

Here we present laboratory dusty plasma experiments in which either a {\it static} or a {\it periodic} shear is applied on a single layer dusty plasma in the strongly coupled regime. During the evaluation of the experimental data we adopt an earlier experimental method \cite{NosenkoPRL04} as well as specific methods used so far only in molecular dynamics (MD) simulations. Our experimental results are supported by, and are combined with simulations and theoretical calculations. To obtain the most complete information about the complex plasma single layer we perform three different experiments on the same dust cloud: (1) analysis of the thermally excited waves, without any applied shear, to determine the principal system parameters, (2) applying a static shear ($\omega_{\rm sh}=0$) to investigate stationary flows at high shear rates, and (3) applying a periodic shear ($\omega_{\rm sh}>0$) to obtain the frequency-dependent complex viscosity.

The paper is structured as follows. Section II describes the details of the experimental apparatus. Section III discusses our experiments and the data evaluation, as well as the connection with simulation data. Section IV summarizes the work.

\section{Experimental apparatus}

Our dusty plasma experiments are carried out in a custom designed vacuum chamber with an inner diameter of 25~cm and a height of 18~cm. The lower, powered, 18~cm diameter, flat, horizontal, stainless steel electrode faces the upper, ring shaped, grounded aluminum electrode, which has an inner diameter of 15~cm and is positioned at a height of 13~cm. 

The experiments are performed in an argon gas discharge at a pressure $p = 1.2 \pm 0.05$~Pa, and at a steady gas flow of $\sim 0.01$~sccm. We apply 13.56~MHz radio frequency excitation of $\sim 7$~W power, to establish the discharge in which the melamine-formaldehyde micro-spheres with a diameter $d = 4.38 \pm 0.06~{\rm \mu m}$ and a mass $m=6.64\times10^{-14}$~kg are levitated. For illumination of the particle layer we use a 200~mW, 532~nm laser, the light of which is expanded and enters the chamber from a side window. Our CCD camera has a resolution of 1.4 Megapixels and captures snapshots of the particle ensemble through the top window of the chamber at 29.54 frames per second acquisition rate. The camera is sensitive only at the illuminating laser wavelength, due to the application of an interference filter centered at 532~nm wavelength. 

The average dust particle number in the field of view is $\sim 2500$, while the total particle number in the dust cloud is about $\sim 15000$. During the evaluation of the raw images identification and position measurement of the particles is performed using the method described in \cite{Feng07}.  

\begin{figure}[tb]
\includegraphics[width=\columnwidth]{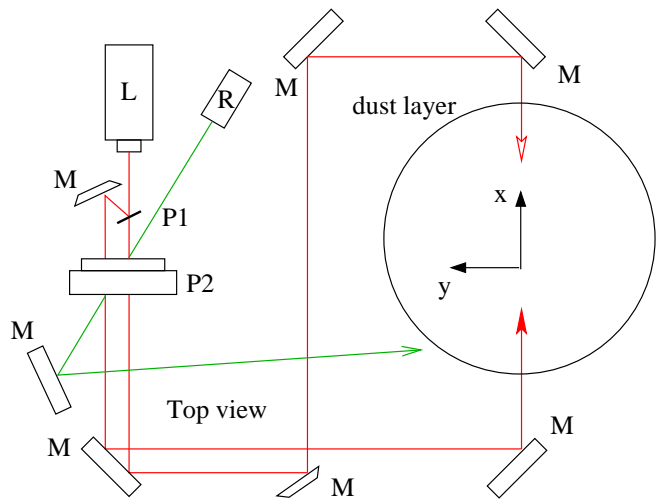}
\caption{\label{fig:opt} 
(color online) Scheme of the optical setup used to generate the alternating shearing laser beams. Main parts: L - 200~mW 650~nm diode laser; P1 - polarizing beamsplitter; M - mirror; P2 - linear polar filter mounted on a ball bearing and rotated by a motor. Beams with different arrow styles are modulated antiphase. In the reference beam line: R - 10~mW 532~nm diode laser, M - mirror.}
\end{figure}

The scheme of the optical setup used in the measurements to apply an external shear on the dust layer, is shown in Fig.~\ref{fig:opt}. The light of a 200 mW red diode laser (L) is split into two parallel beams of equal intensity and perpendicular linear polarization. One of these beams is directed towards another polarizer (P2), while the beam with the perpendicular polarization reaches P2 after being reflected from a mirror. By rotating P2 by means of a motor the intensities of these two beams, passing through P2, can be modulated harmonically, with $180^\circ$ phase shift relative to each other. These beams are further guided by additional mirrors, which are aligned to ensure that the two beams propagate inside the chamber in opposite directions, but on a common axis. This axis is adjusted to be horizontal and its vertical position is set in a way that the beams, which are focused to have a diameter of about 0.5~mm, lie within the dust layer.

This setup makes it possible to exert a periodic force on the particles, which, in the center of the beam(s) has a form $F_x(t) = F_0\sin(\omega_\text{sh} t)$, with the direction $x$ defined to coalesce with the beams, as shown in Fig.~\ref{fig:opt}. The amplitude $F_0$ can be tuned by the power of the laser (L), while the frequency $\omega_\text{sh} = 2\pi / T$ is defined by the half-period of rotation ($T$) of P2. As the recording of the images by the camera is not synchronized with the rotation of the polarizer, the angular position of the latter has to be measured. This is accomplished by a reference beam of a green diode laser (R). The polarized beam of this laser passes through P2 as well, and when P2 is rotated, the beam intensity is modulated. The beam is directed onto the surface of the lower electrode of the discharge, and, as the wavelength is the same as that of the laser illuminating the particle layer, the beam spot is picked up by the camera on the images of the particle layer. Measuring the intensity of this reference beam in the recorded images allows an accurate determination of the angular position of P2.

In the experiments aimed at the study of the static viscosity P2 is not rotated, it is aligned to a position, which maximizes the intensity of one of the beams.

\section{Experiments}

This section is divided into three parts. Section III.A explains the ``calibration'' of the system, which was aimed at the determination of the basic characteristics of the dust layer (particle charge, screening parameter of the potential, plasma frequency), needed in the subsequent data analysis. Section III.B discusses the measurements of the static viscosity covering a wide domain of applied shear rates. These measurements quantify shear thinning at elevated shear rates. A brief introduction of the simulation method is presented in this section. Section III.C describes the experiments carried out with a harmonic perturbation, which reveal the viscoelastic behavior of the investigated system.

\subsection{Calibration}

The aim of this experiment is the determination of the basic parameters of our particle suspension, which will be crucial in the forthcoming experiments aimed at the measurement of viscous properties. Having calibrated the optical system the Wigner-Seitz radius is determined from the average particle separation. We find an areal density $n=3.21$~mm$^{-2}$ and the Wigner-Seitz radius to be $a=1/\sqrt{\pi n}=0.315$~mm. Fitting the particle velocity distribution with a Maxwellian distribution resulted in a mean thermal velocity $v_{\rm th,0} = \sqrt{2kT/m} = 0.66$~mm/s, for the unperturbed system.

\begin{figure}[htb]
\includegraphics[width=1.0\columnwidth]{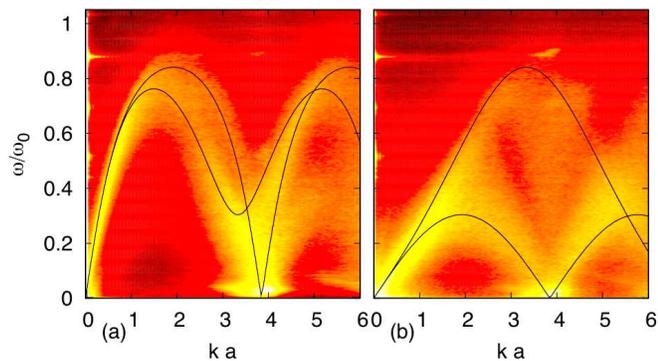}
\caption{\label{fig:disp} 
(color online) Longitudinal (a) and transverse (b) current fluctuation spectra from the experiment (color / grayscale map) with overlaid Yukawa ($\kappa=1.2$) lattice dispersion curves for the two principal lattice directions (lines).}
\end{figure}

The determination of the parameters characterizing the interparticle interaction (particle charge, Debye screening length, plasma frequency) is based on the dependence of  the longitudinal and transverse wave dispersion relations on these quantities, as introduced in \cite{Nunomura2002,*Zhdanov2003}. Via tracing the trajectories of the particles we calculate the longitudinal and transverse current fluctuation spectra from the ${\bf r}_i(t)$ position and ${\bf v}_i(t)$ velocity data as described in \cite{DonkoJPC08}. For our system we adopt the Debye-H\"uckel (Yukawa) type interaction potential, giving a pair interaction energy: $U(r) = Q\Phi(r)$. Matching the dispersion relations obtained from these spectra with theoretical dispersion curves of Yukawa lattices \cite{lattice} (see Fig.~\ref{fig:disp}) we have obtained the following system parameters (within an uncertainty of $\pm 8$ \%): Debye screening length $\lambda_D=0.263$~mm, screening parameter $\kappa=a/\lambda_D=1.2$, particle charge $Q = 4840 e$ (where $e$ is the electron charge), and nominal plasma frequency $\omega_0=\sqrt{n Q^2/2\varepsilon_0 m a} = 72.1$~rad/s.

\subsection{Experiment with static shear}

As explained in Section II, in this experiment the polarizer P2 (see Fig.~\ref{fig:opt}) is not rotated, it is set in a way that one of the laser beams carries the full power. Two methods are available to obtain viscosity data from this type of experiments. 

Method 1 is based on the solution of the Navier-Stokes (NS) equation. In \cite{NosenkoPRL04} this method was applied for a system sheared by two (horizontally shifted), parallel, counter-propagating laser beams. We apply the same approach for our system sheared by a single beam. In this case the NS equation results in a velocity profile $v_x(y) = v_0 \exp(-\sqrt{\nu \rho / \eta}~|y|)$, where $v_0$ is the stationary flow velocity in the beam center, $\nu$ is the effective dust-background collision frequency (frictional drag), and $\rho$ is the 2D particle mass density. The advantage of this method is, that it is relatively noise insensitive and does not assume any particular form for the interaction pair potential. The drawbacks are that (i) one obtains spatially averaged data, and (ii) only for the ratio of viscosity and collision frequency, thus an estimation or independent measurement of $\nu$ is needed, which can introduce additional uncertainty. 

\begin{figure}[htb]
\includegraphics[width=\columnwidth]{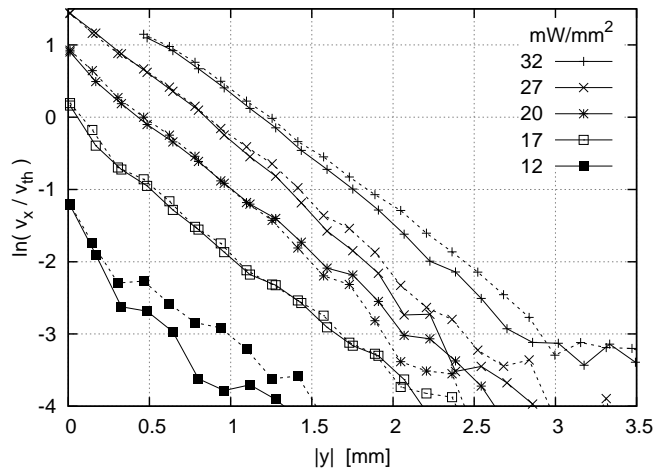}
\caption{\label{fig:vel} 
$v_x(y)$ velocity profiles for different shearing laser power densities in the dust particle layer in units of the average thermal velocity $v_\text{th} = 1.8$~mm/s. Different lines belong to the two sides ($y>0$ and $y<0$) of the sheared region.}
\end{figure}

Figure~\ref{fig:vel} shows time averaged experimental $v_x(y)$ velocity profiles for different shearing laser powers. A $\nu/\eta$ value of $1.64\times10^{13}$~kg$^{-1}$ is derived from these data. The average thermal velocity, estimated form the average peculiar velocities of particles (defined as the velocities relative to the flow) within the sheared region, was found to lie between $v_{\rm th} = 1.6$~mm/s and 2.0~mm/s for all investigated cases using different shearing laser powers, about three times higher compared to the unperturbed case. This increase is due to the energy absorbed from the shearing laser beam. This effect is called shear induced melting and was studied for dusty plasma crystals in detail in \cite{Nosenko09,GoreePRL10b}. As a result of this heating process our system is in the strongly coupled liquid phase within the sheared region. In the following we use the average value $v_{\rm th} = 1.8$~mm/s for the presentation of our results.

Method 2 of data acquisition was used so far in molecular dynamics simulations only \cite{DonkoPRL06,Ramazanov,DonkoPRE10} and is based on the measurement of the applied shear stress, or, equivalently the off-diagonal element of the pressure tensor:
\begin{equation}\label{eq:P}
P_{xy} = \frac{1}{A}\bigg[ \sum_i m v_{i,x} v_{i,y} + \frac{1}{2}\sum_i\sum_{j\ne i} r_{ij,y} F_{ij,x}\bigg],
\end{equation}
where $A$ is the area of the region of interest, $v_{i,x}$ is the $x$ component of the peculiar velocity of particle $i$, $r_{ij,y}$ is the $y$ component of the distance vector between particle $i$ and $j$, $F_{ij,x}$ is the $x$ component of the force acting on particle $i$ due to the pair interaction with particle $j$. Summation of $i$ is for particles within $A$, while summation of $j$ runs over all particles interacting with particle $i$. The viscosity is obtained by calculating
\begin{equation}\label{eq:eta}
\eta =   \frac{-P_{xy}}{\dot{\gamma}}.
\end{equation} 

\begin{figure}[htb]
\includegraphics[width=\columnwidth]{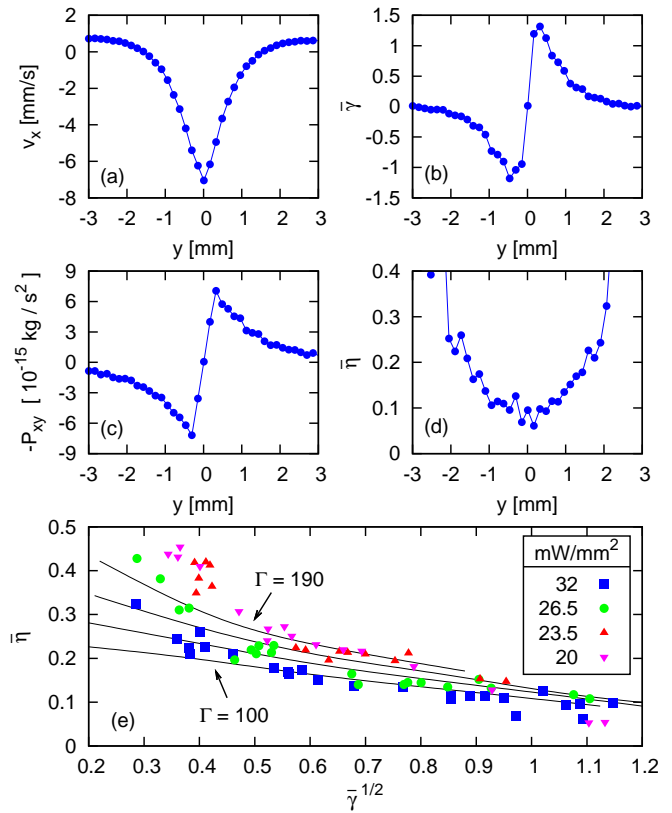}
\caption{\label{fig:static} 
(color online) Spatial dependence of the (a) velocity profile $v_x(y)$; (b) normalized shear rate $\bar{\gamma}$, ; (c) off-diagonal element of the pressure tensor $-P_{xy}$; (d) normalized viscosity. (e): shear rate dependence of the viscosity, $\bar{\eta}(\bar{\gamma}^{1/2})$, for different laser power densities. The solid lines show MD simulation results for $\Gamma$ values indicated.}
\end{figure}

In our experiment the shear rate strongly depends on the $y$-coordinate, due to the exponential velocity profile shown in Fig.~\ref{fig:vel}. For further evaluation we set up a computational grid along the $y$-axis with a resolution of $\Delta y = 0.16$~mm. In this way $P_{xy}(y)$ and $\dot{\gamma}(y)$ can be measured by reducing the evaluation volume [$A$ and the summation of $i$ in (\ref{eq:P})] to one grid interval at a time. Intermediate results obtained with 32 mW/mm$^2$ laser power density shear are shown in Fig.~\ref{fig:static}(a-d), illustrating the evaluation process: the observed and time-averaged velocity profile (a) is used to compute its derivative, the shear rate, which is given in (b) in normalized units, $\bar{\gamma} = (\partial v_x/\partial y)(a/v_{\rm th})$ (with $v_{\rm th} = 1.8$~mm/s). The spatial distribution of the off-diagonal element of the pressure tensor follows the trend of the shear rate (c). The viscosity is the ratio of the data in (c) and (b), and is shown in (d). Substituting the $y$-coordinate with the shear rate taken from (b) results in the {\it shear rate dependence} of the viscosity, shown in (e) in normalized units, $\bar{\eta} = \eta/\eta_0$, where $\eta_0=mn\omega_0 a^2 = 1.52\times 10^{-12}$~kg/s.    

We have adapted our molecular dynamics (MD) simulation code \cite{DonkoPRL06,*DonkoMPL07,DonkoPRE10} to the experimental conditions. This non-equilibrium MD method is based on the homogeneous shear algorithm \cite{Evans}, which has frequently been used in studies of the viscosity of liquids. The approach is based on the Gaussian thermostated SLLOD equations of motion and applies the Lees-Edwards sliding periodic boundary conditions (for details, see \cite{Evans}). The shear rate, which can be stationary or oscillatory, is an input parameter, the off-diagonal element of the pressure tensor is computed based on (\ref{eq:P}), where the summation volume $A$ is the whole simulation box. Viscosity is than calculated using (\ref{eq:eta}), as it is done for the experiment.

The simulation results are shown in Fig.~\ref{fig:static}(e). An important difference between the experiment and the simulation is in the working principle of the thermostat. In the simulation a linear velocity profile is achieved together with a uniform temperature distribution in the whole simulation cell \cite{Evans}. In this case the system can be well characterized with the Coulomb coupling parameter $\Gamma$. In the experiment the temperature is not uniform, the system is in crystalline state outside the sheared region and is in liquid state inside. An estimation of an average $\langle\Gamma\rangle \approx 130$ can be given based on the $v_{\rm th}$ data averaged over the investigated region in space. This result confirms non-Newtonian behavior \cite{DonkoPRL06}, the so called shear-thinning effect, where the viscosity decreases with increasing shear rate. 

In earlier experiments, like \cite{LinI2001,NosenkoPRL04,Gavrikov05}, a variation of the shear viscosity with changing laser intensity was already observed, however, it was attributed to the temperature variation near the shearing laser beam(s). The agreement between our spatially resolved experimental results and our fully thermostated simulations points out, that the effect of temperature in this range is marginal, it is the high shear rate that causes most of the variations of the shear viscosity.

Taking the average viscosity value of $\bar{\eta}\approx 0.15$ ($\eta\approx 2.2\times 10^{-13}$~kg/s) for intermediate shear rates and comparing it to the ratio $\nu/\eta = 1.64\times10^{13}$~kg$^{-1}$, as obtained when applying Method 1, we estimate the effective dust-background collision frequency to be $\nu \approx 3.7$~s$^{-1}$. This value is slightly higher than the approximated dust-neutral collision frequency defined as $\nu_{\rm dn}=\delta(4\pi/3) r_d^2 n_n v_n m_n/m_d \approx 3.1$~s$^{-1}$, where $r_d$ and $m_d$ are the dust radius and mass; $n_n$, $v_n$ and $m_n$ are the neutral number density, average speed and mass, respectively, $\delta=1.26$ \cite{Liu2003}.

\subsection{Experiment with harmonic shear}

Turning on the rotation of the polar filter P2 (see Fig.~\ref{fig:opt}) results in a sinusoidally modulated periodic shear in the $x$ direction. The amplitude of the modulated laser power density is set to 32 mW/mm$^2$. The frequency is varied between $\omega_{\rm sh}=3.7$ and 44~rad/s in 25 steps. 

\begin{figure}[tb]
\includegraphics[width=\columnwidth]{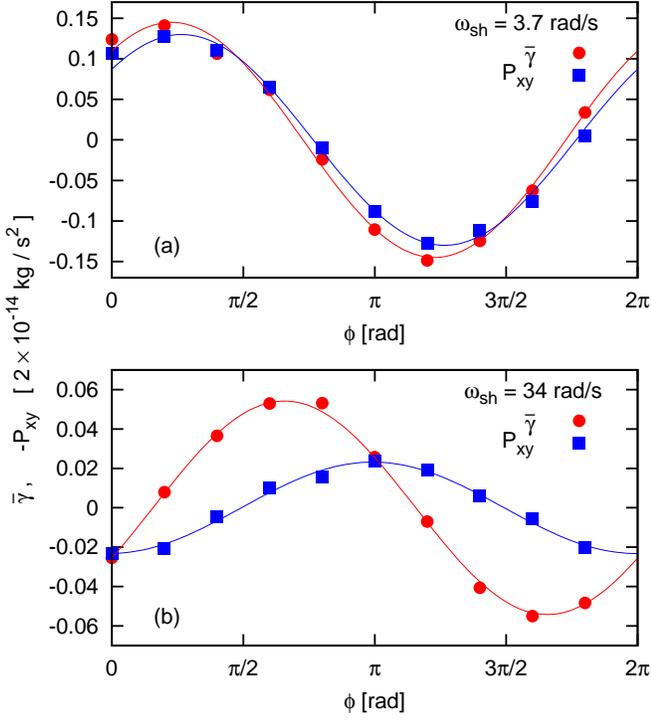}
\caption{\label{fig:fig5} 
(color online) Measured phase-resolved shear rate ($\bar{\gamma}$) and off-diagonal element of the pressure tensor ($-P_{xy}$) for a low (a) and a high (b) frequency periodic shear (symbols) and least-square fitted sine functions (lines). }
\end{figure}

To optimize the signal to noise ratio phase resolved averaging is performed: individual oscillation cycles are identified based on the reference signal (as explained in section II). The period time is divided into ten time slots, and every snapshot is assigned to one of them according to its actual phase angle $\phi$ relative to the reference signal. Despite the low number of particles in narrow slices in space and time, this type of averaging $\sim 1800$ snapshots at a given frequency makes the evaluation Method 2 (based on the pressure tensor and shear rate) possible. Further, $\dot{\gamma}(\phi,y)$ and $P_{xy}(\phi,y)$ are averaged in space in the sheared region $0.5~\text{mm}<|y|<1.5~\text{mm}$ (see Fig.~\ref{fig:static}). Examples of the measured $\dot{\gamma}(\phi)$ and $P_{xy}(\phi)$ for selected frequencies are presented in Fig.~\ref{fig:fig5} together with least squares fits in the form $f(\phi)=\xi\sin(\phi+\phi_0)$. 

Performing the sine function least-square fitting procedure we obtain the amplitude and phase for both the pressure and shear rate for each excitation frequency: $\xi^P(\omega_{\rm sh})$, $\xi^\gamma(\omega_{\rm sh})$, $\phi_0^P(\omega_{\rm sh})$, and $\phi_0^\gamma(\omega_{\rm sh})$, respectively. 

\begin{figure}[tb]
\includegraphics[width=\columnwidth]{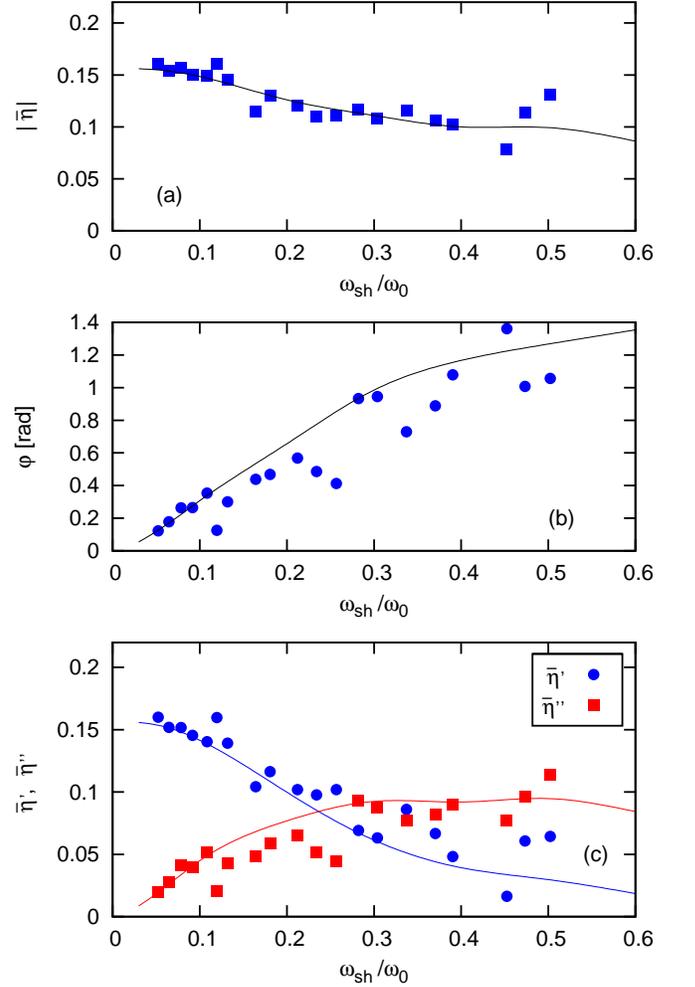}
\caption{\label{fig:dyn1} 
(color online) Frequency dependence of the shear viscosity: (a) magnitude $|\eta(\omega)|$, (b) complex argument $\varphi(\omega)$, and (c) the real and imaginary parts $\eta^{\prime}(\omega)$ and $\eta^{\prime\prime}(\omega)$. The lines show MD simulation results for $\Gamma = 200$.}
\end{figure}

The magnitude of the frequency dependent shear viscosity can be calculated from the amplitudes, while the complex argument is given by the difference of the phases:
\begin{eqnarray}
|\eta(\omega_{\rm sh})| &=& \frac{\xi^P(\omega_{\rm sh})}{\xi^\gamma(\omega_{\rm sh})},\\ \varphi(\omega_{\rm sh}) &=& \phi_0^P(\omega_{\rm sh}) - \phi_0^\gamma(\omega_{\rm sh}). \nonumber
\end{eqnarray}
as shown in fig.~\ref{fig:dyn1}(a) and (b).

The real and imaginary (dissipative and elastic) parts of the complex viscosity $\eta(\omega_{\rm sh}) = \eta^\prime(\omega_{\rm sh}) - i\eta^{\prime\prime}(\omega_{\rm sh})$ are computed using the magnitude and the complex argument as 
\begin{eqnarray}
\eta^\prime(\omega_{\rm sh}) &=& |\eta(\omega_{\rm sh})|\cos\left[\varphi(\omega_{\rm sh})\right],\\ \eta^{\prime\prime}(\omega_{\rm sh}) &=& |\eta(\omega_{\rm sh})|\sin\left[\varphi(\omega_{\rm sh})\right].\nonumber
\end{eqnarray}

With increasing $\omega_\text{sh}$ we observe a decrease of $\eta^\prime$ and an increase of $\eta^{\prime\prime}$, as shown in Fig.~\ref{fig:dyn1}(c). The crossover of the real and imaginary parts can be observed at a frequency $\omega_{\rm cross}/\omega_0 = 0.3 \pm 0.05$. Our MD simulation (using input parameters obtained from the experiment) show remarkably good agreement with the experimental data.

\section{Summary}

We have studied the static and dynamic shear viscosity of a complex plasma layer via the combination of different experimental and simulation techniques. We have developed an optical setup by means of which we applied a static, as well as a harmonically modulated shear on the particle suspension. 

In the experiments using static shear the viscosity was measured via the calculation of the pressure tensor elements form the recorded particle trajectories. The measurements have quantified the shear thinning effect, the decrease of the shear viscosity with increasing shear rate. These results are consistent with an earlier experiment \cite{NosenkoPRL04}. Molecular dynamics simulations, modeling a homogeneously sheared liquid with system parameters taken from the experiment support our findings and provide a possibility to rule out the dominance of thermal effects on the observed variations of the shear viscosity.

In the experiments with sinusoidally modulated shear the pressure tensor elements have been measured in a time-resolved manner. This way we were able to determine the viscoelastic response of the dust layer. The data have indicated a phase delay between the perturbation and the response, that increases with increasing frequency: we observed a gradual decrease of the viscous term (real part of the viscosity) and a gradual increase of the elastic term (imaginary part of the viscosity). These experimental results are very similar to those found for 3D Yukawa liquids in molecular simulations \cite{DonkoPRE10}.

The appearance of such a rich spectrum of non-Newtonian behavior observed in complex plasma experiments can perhaps be explained by referring to the manyfold of interactions and effects between the dust and the background plasma. Similarly complex plastic response is observed in molecular fluids, like polymer or organic solutions, paints, etc. The fact that our molecular dynamics simulations, considering an idealized Yukawa system (i.e. neglecting friction, ion wakes, etc.), well reproduce the viscoelastic properties of the experimental system, however raises new questions. First, what microscopic feature is minimally necessary to result in a non-Newtonian macroscopic response? Second, at which point does the real complexity of dusty plasmas introduce qualitatively new macroscopic features, not reproducible by the simplified numerical models based on isotropic Yukawa interaction? We can point out only two features, namely, the long range microscopic interaction and the strongly coupled (correlated) macroscopic state of the model system that may in some way be responsible for the computationally observed non-linearities.

\begin{acknowledgments}
This research has been supported by the Grants OTKA PD-75113, K-77653, and the J\'anos Bolyai Research Foundation of the Hungarian Academy of Sciences.
\end{acknowledgments}


%

\end{document}